\documentclass[11pt, a4paper]{article}
\usepackage[utf8]{inputenc} 
\usepackage[margin=1in]{geometry} 
\usepackage{amsmath, amssymb} 
\usepackage{graphicx} 
\usepackage{booktabs} 
\usepackage{hyperref} 
\usepackage{microtype} 
\usepackage{caption}
\usepackage[utf8]{inputenc}
\usepackage[margin=1in]{geometry}
\usepackage{amsfonts, amssymb, amsmath}
\usepackage[numbers, sort&compress]{natbib}
\hypersetup{
    colorlinks=true,
    linkcolor=blue,
    filecolor=magenta,
    urlcolor=cyan,
    citecolor=blue,
}
\title{
  \vspace{-2em}
  \hrule height 2pt 
  \vspace{0.5em}
  {\LARGE\bfseries The CRAFT principles for the responsible use of large language models in policymaking}
  \vspace{0.5em}
  \hrule height 0.5pt 
  \vspace{1em}
}
\author{
  \textbf{Willem Fourie}$^{1,}$\thanks{Corresponding author. Email: \texttt{willemf@sun.ac.za}} \quad
  \textbf{Gray Manicom}$^{1}$ \quad
  \textbf{Tanya de Villiers-Botha}$^{2}$
  \\[1em]
  \small $^{1}$School for Data Science and Computational Thinking, Stellenbosch University \\
  \small $^{2}$Department of Philosophy, Stellenbosch University
}
\date{} 
\begin{document}
\maketitle

\begin{abstract}
Policymakers around the world face the question of how to use artificial intelligence in general, and large language models in particular, to improve the policymaking process. Used well, large language models can strengthen the collection, interpretation and synthesis of policy-relevant information and the drafting of policy-relevant output. Yet the use of large language models in policymaking is associated with risks. Output that is plausible but not necessarily correct, bias resulting from unrepresentative training data, the exposure of sensitive information and, over time, deskilling and dependency can erode trust if large language models are not used thoughtfully. The CRAFT principles -- control, rigour, accountability, fairness and transparency -- offer a way to make the most of large language models in policymaking while managing the risks.
\end{abstract}
\vspace*{0.4em}

\section{Understanding large language models}

\subsection{What is a large language model?}

Artificial intelligence (AI) is the field of building machines that simulate aspects of human intelligence. Machine learning is one approach within the field of AI, where machines learn patterns from data. Neural networks are a family of machine learning methods, loosely modelled on the human brain, that improve as more data and computing power become available. A large language model (LLM) uses an architecture called the transformer, which consists of neural network components, and learns linguistic relationships by being trained on large quantities of text \citep{vaswani2017}. LLMs learn to predict the next word in a sequence. Many LLMs exist, including ChatGPT, Gemini, Claude, DeepSeek and Llama.

\subsection{How are large language models trained?}

Training LLMs typically takes place in three broad stages \citep{ouyang2022, mitchell2024}:

\begin{itemize}
  \item During the pre-training stage, the model is trained on a very large body of text, mostly taken from the internet. By repeatedly predicting next or missing words across billions or trillions of examples, the model learns the structure of language and acquires its general capabilities.
  \item During the fine-tuning stage, the model is further trained on smaller datasets to improve its performance on specific tasks. This can improve its domain knowledge, linguistic style or ability to follow instructions.
  \item During the alignment stage, the model is adjusted towards human preferences and safety requirements.
\end{itemize}

Further limits can also be imposed at deployment, through specific instructions, human review and filters.

\subsection{How do large language models produce their output?}

LLMs generate text one token at a time. Tokens are common word fragments or symbols, rather than whole words. For a given input sentence, the LLM returns a set of probabilities across all possible next tokens in its language model, one token is chosen from that set and added to the sentence, and the process is repeated until the response is complete. Those probabilities are calculated in the following way \citep{bengio2008, mikolov2013, mitchell2026}:

\begin{itemize}
  \item The input text is split into tokens.
  \item Tokens are converted into number vectors to be analysed mathematically. These vectors represent context-dependent concepts as directions in a geometric space in which related concepts align.
  \item The model processes all the tokens simultaneously, which provides context, and outputs a probability for every possible next token.
  \item One token is sampled from that distribution, with different LLMs using different sampling strategies. This means the same input will not necessarily produce the same output.
\end{itemize}

The probabilities an LLM calculates derive from the text it was trained on. Therefore, what is absent from the training data is largely absent from what the model produces. It also implies that the model will be more likely to produce text completions that commonly appear in the training data or that have the same semantic meaning. An LLM's outputs, therefore, reflect its training data and training process, not necessarily reality as humans experience it \citep{bommasani2021}.

\subsection{What a large language model is not}

When reflecting on the responsible use of LLMs in policymaking, it is tempting to understand them in human terms, and thus to anthropomorphise these technologies. Doing so is a category error: although an LLM reproduces the patterns of human communication, it is qualitatively different from a human and should be treated as such \citep{mitchell2026}.

It is more useful to view an LLM as analogous to a calculator, with one qualification: where a calculator returns an exact, correct number, an LLM returns a plausible continuation of a text. Like a calculator, an LLM is a tool that performs a function without any understanding of what it is doing.

In policymaking contexts this has a direct consequence for the principles of responsible LLM use. An LLM is not an author, and using one without disclosure is therefore not plagiarism, since plagiarism is the appropriation of another person's work and there is no such person behind a model's output. It means the obligation to disclose AI use rests on different grounds, and the disclosure of LLM use should be proportionate to what it was used for.

\section{Large language models and policymaking}

\subsection{What is policymaking?}

There are various ways to understand policymaking, including as a process that moves through distinct stages \citep{lasswell1956, howlett2009}, as `muddling through' under conditions of bounded rationality \citep{simon1955, lindblom1959} or as competition between advocacy coalitions held together by shared belief systems within policy subsystems \citep{sabatier1993}.

We frame policymaking as information processing \citep{jones2005}. From this perspective, the attention of policymakers and institutions is a scarce resource, and prioritisation becomes a defining feature of policymaking amid an oversupply of information \citep{workman2009}. We assume that none of the components of information processing is neutral \citep{stone2002}.

Within this framing, policymaking covers four functions:

\begin{itemize}
  \item the collection of relevant, representative and reliable information;
  \item the rigorous interpretation of information;
  \item the transparent and balanced synthesis of information; and
  \item the drafting of clear and legitimate policies.
\end{itemize}

These four functions ultimately contribute towards policies that are accepted as legitimate and enable positive societal impacts.

\subsection{How can large language models support policymaking?}

LLMs have the potential to significantly enhance each of the functions of policymaking:

\begin{itemize}
  \item \textit{Collection:} LLMs can search more text than a team could read. They can find or collect relevant material across large bodies of text, such as citizen views, existing policies, parliamentary debates, quantitative data and social and conventional media.
  \item \textit{Interpretation:} LLMs can provide human-reviewable comparisons and analyses of large amounts of information. Done well, this capability can deepen human analysis of large and diverse datasets, previously difficult to navigate.
  \item \textit{Synthesis:} LLMs can surface human-reviewable patterns across more documents than a team could read and produce a first consolidated view for experts to interrogate. Conversely, LLMs can be used to interrogate human interpretations, thereby identifying potential human biases.
  \item \textit{Drafting:} LLMs are currently perhaps the most capable as drafters in policymaking settings. Used well, LLMs can produce fluent first drafts, restructure unclear text and render the same content in different registers for different audiences.
\end{itemize}

\subsection{What are the risks when using large language models in policymaking?}

Various risks accompany the use of LLMs in policymaking. The most widely discussed is inaccuracy \citep{huang2025}. Because LLMs generate confident, fluent and plausible text rather than verified facts, the errors they produce, including wholly fabricated sources and figures, can be difficult to detect \citep[e.g.][]{kalai2025}.

Closely related, and also well researched, is bias \citep[e.g.][]{gu2019, gupta2024, mehrabi2021}. The training data underlying many commercial models is concentrated in a small number of languages and countries. As a result, these systems can reflect and reinforce biases against people, communities and countries that are under-represented in the data and perform worse in those contexts.

Where inaccuracy and bias relate to what a model produces, a further risk concerns information provided to an LLM \citep{das2025, yan2024}. Information entered into a model may be stored, retained or used to train future systems. Consequently, personal or sensitive material that ought to remain protected can be exposed or misused. Privacy therefore requires careful consideration before such information is entered into a model at all.

Other risks relate less to any single output than to the cumulative effects of relying on the technology over time. The danger of deskilling is one such risk \citep{macnamara2024, ferdman2026, novelli2026}. As policymakers rely on LLMs to gather information, conduct analysis or draft documents, their own analytical and drafting capabilities may weaken, both individually and institutionally. A related risk is dependency \citep{bilderback2025, budhwar2023}. As LLM-based tools become embedded in everyday workflows, often enabled by default and governed by terms that can change without notice, policymakers may come to rely, sometimes unknowingly, on systems and infrastructure they do not control. This can reduce their autonomy and erode sovereignty over the data, processes and capabilities on which policymaking depends.

\section{CRAFT: principles for using large language models in policymaking}

Used appropriately, large language models can strengthen the policymaking process. Their benefits, however, do not arise by default. The CRAFT principles set out the conditions under which these tools can be used responsibly in policymaking.

The table below summarises the five principles, which are then discussed in turn.

\begin{table}[ht]
  \centering
  \caption{The CRAFT principles for the responsible use of large language models in policymaking.}
  \label{tab:craft}
  \begin{tabular}{@{}p{2.3cm} p{6.0cm} p{6.0cm}@{}}
    \toprule
    \textbf{Principle} & \textbf{Description} & \textbf{Practice} \\
    \midrule
    Control & People and institutions stay in control of the work and the tools, and remain able to do the work without the model. & Decide what the LLM will be used for, keep the capacity to work without it, and choose the model, data and infrastructure rather than accepting a provider's defaults. \\
    \addlinespace
    Rigour & An LLM's output is plausible but not necessarily correct, and should be treated as unverified until checked. & Verify LLM-derived information against original sources and test it through human expertise, drawing on specialists where needed. \\
    \addlinespace
    Accountability & A named person answers for the work and cannot pass that responsibility to the model. & Assign a specific person as answerable for each decision the LLM contributes to. \\
    \addlinespace
    Fairness & Under-representation in an LLM must not carry through into the policy: the relevant groups, ideas and perspectives must be fairly represented. & At each stage, identify who or what the model has under-represented or misrepresented and draw on relevant sources from beyond the model. \\
    \addlinespace
    Transparency & An LLM's use is disclosed so the legitimacy of the resulting policy is preserved. & State where and how the LLM was used, with disclosure proportionate to its role and fuller as that role moves from expression to substance. \\
    \bottomrule
  \end{tabular}
\end{table}

\subsection*{Control}

Policymakers must remain in control of how an LLM is used. This is the governing principle across all four policymaking functions and includes the choice of whether to use an LLM at all as well as on what information and on whose infrastructure the LLM is used. These decisions should be reviewed regularly.

Practically, this means deciding what the LLM is used for, keeping the capacity to perform the work without it, and choosing the model, the data and the infrastructure deliberately rather than accepting what a provider supplies by default.

Application of this principle should allow policymakers to address both individual and institutional risks. For individual policymakers, it supports the use of LLMs that enhance rather than erode their expertise. Institutionally, it should similarly strengthen rather than weaken institutions, by ensuring the locus of control remains within policymaking institutions.

\subsection*{Rigour}

An LLM's output seems plausible but is not necessarily correct, and should therefore be treated as unverified until it is checked. Rigour is particularly relevant for the interpretation of information and assessing the reliability of what is collected. The breadth, analytical quality and clarity that an LLM can offer ultimately depend on this rigour.

Practically, this means that information identified by or derived from an LLM should be verified against original sources and tested through human expertise and judgement, drawing on additional specialist input where necessary.

Application of this principle addresses potential inaccuracies of LLM outputs in each of the four policymaking functions. At the same time, it positions the LLM as an enhancement of human expertise rather than a replacement for it.

\subsection*{Accountability}

A named person must be able to answer for the work an LLM has contributed to, and that responsibility cannot be transferred to the LLM. Accountability is relevant in all four functions of policymaking -- spanning from which sources are used, through how they are interpreted and combined, to how the result is expressed -- and becomes possible only where control and rigour are already in place.

Practically, this means that a specific person should be answerable for each decision the LLM contributed to.

Application of this principle allows those affected by policies to exercise their right to hold policymakers to account, and should provide an entry point to improving policies through further societal input.

\subsection*{Fairness}

Under-representation and bias in an LLM should not carry through into policy. Because an LLM reflects its training data, and because that data under-represents much of the world, specific emphasis should be placed on ensuring fair representation of the relevant groups and perspectives.

Practically, this means identifying, when collecting, interpreting, synthesising and drafting, who or what the LLM is likely to have under-represented, and deliberately drawing relevant additional information in from sources beyond the LLM.

Application of this principle directly addresses the bias inherent to any LLM. It also broadens participation in the policymaking process by allowing those affected by a policy to highlight the potential exclusion of groups and perspectives.

\subsection*{Transparency}

An LLM's use should be disclosed in a way that strengthens the legitimacy of the resulting policy, which in turn depends on people being able to understand how a policy was drafted.

Practically, this means stating where and how an LLM was used. The extent of that disclosure should be proportionate to the LLM's role, increasing as the role moves from improving the language and grammar of a text to contribution to its substance.

Application of this principle makes the use of an LLM visible to those affected by a policy. This visibility supports the legitimacy of the policy, but only where control, rigour, accountability and fairness have already been applied.

Taken together, the CRAFT principles could enable policymakers to make the most of the benefits of large language models while managing the risks. Applied consistently and reviewed as the technology develops, they could contribute to ensuring that these tools strengthen the policymaking process rather than erode it.

\bigskip
\noindent{\small\textbf{Disclosure of AI use.} A large language model was used to assist in preparing this document. It supported the collection and synthesis of source material, drawing on expert input from a Stellenbosch University webinar on the responsible use of LLMs in government, combined with an additional analysis and integration of policy and technical literature. The LLM was also used in critiquing key components of the document and provided the first draft of the summary table. The LLM was further used in the final editing of the brief. The substantive expertise and the argument are the authors'. All factual claims and sources were verified, and the authors take responsibility for its content.\par}

\bibliographystyle{plainnat}
\bibliography{references}

\end{document}